\documentclass[twocolumn,amsmath,amssymb,aps]{revtex4-1} 
\usepackage[utf8x]{inputenc}
\usepackage{palatino}

\usepackage{graphicx}
\usepackage[caption=false]{subfig} 
\usepackage{bm}

\newcommand{\ab}{{\alpha\beta}}

\long\def\comment#1{}

\begin{document}

\title{Probable nature of higher-dimensional symmetries \\underlying mammalian grid-cell activity patterns} 

\author{Alexander Mathis}\email{amathis@fas.harvard.edu}\affiliation{Department of Molecular and Cellular Biology, and Center for 
Brain Science, Harvard University, Cambridge, Massachusetts 02138, USA}

\author{Martin B. Stemmler} \author{Andreas V.M. Herz}\affiliation{Bernstein Center for Computational 
Neuroscience Munich and Fakult\"at f\"ur Biologie, Ludwig-Maximilians-Universit\"at M\"unchen,  82152 Martinsried, Germany}
\date{\today}

\begin{abstract}  
\noindent
Lattices abound in nature - from the crystal structure of minerals to the honey-comb organization of ommatidia in the compound eye of insects.
Such regular arrangements provide solutions for optimally dense packings, efficient resource distribution and cryptographic schemes, highlighting 
the importance of lattice theory in mathematics and physics, biology and economics, and computer science and coding theory. 
Do lattices also play a role in how the brain represents information? To answer this question, we focus on higher-dimensional stimulus domains, with particular emphasis on 
neural representations of the physical space explored by an animal.
Using information theory, we ask how to optimize the spatial resolution of neuronal lattice codes. 

We show that the hexagonal activity patterns of ``grid cells" found in the hippocampal formation of mammals navigating on a flat surface lead to the highest spatial resolution in a two-dimensional world. 
For species that move freely in a three-dimensional environment, the firing fields should be arranged along a face-centered cubic (FCC) lattice or a equally dense non-lattice variant thereof known as a hexagonal close packing (HCP).
This quantitative prediction could be tested experimentally in flying bats, arboreal monkeys, or cetaceans. 
More generally, our theoretical results suggest that the brain encodes higher-dimensional sensory or cognitive 
variables with populations of grid-cell-like neurons whose activity patterns exhibit lattice structures at multiple, nested scales.
\end{abstract}

\maketitle

\section*{Introduction} 

In mammals, the neural representation of space rests on at least two classes of neurons.
``Place cells" discharge when an animal is near one particular location in its environment~\cite{O'KeefeJohnDostrovsky1971}.
``Grid cells" are active at multiple locations that span an imaginary hexagonal lattice covering the environment~\cite{Hafting2005}
and  have been found in rats, mice, crawling bats and human beings~\cite{Hafting2005}\nocite{Fyhn2008}\nocite{Yartsev2011}-\cite{Jacobs2013}. These cells 
are believed to build a metric for space.

In all these experiments, locomotion occurs on a horizontal plane. Information theory shows that the observed hexagonal lattices optimally represent 
such a two-dimensional (2D) space~\cite{Guanella2007b}\nocite{Mathis2012c}-\cite{Wei2013}. 
In general, however, animals move in three dimensions; this is particularly true for birds, tree dwellers, and fish.
Their neuronal representation of 3D space may consist of a mosaic of lower-dimensional patches~\cite{Jeffery2013}, as evidenced by recordings from climbing rats~\cite{Hayman2011}. 
Data from flying bats, on the other hand, demonstrate that their place cells represent 3D space 
in a uniform and nearly isotropic manner~\cite{Yartsev2013}. 

As mammalian grid cells might represent space differently in 3D than in 2D, we study grid-cell
representations in arbitrarily high-dimensional spaces and measure the accuracy
of such representations in a population of neurons with periodic tuning curves. Even though the firing fields between cells overlap, so as to ensure uniform 
coverage of space, we show how resolving the population's Fisher information  
can be mapped onto the problem of packing {\em non-overlapping} spheres.  The optimal lattices are thus the ones with the highest packing ratio--- 
the densest lattices represent space most accurately. 
This remarkably simple and straightforward answer implies that hexagonal lattices are optimal for representing 2D, as others have surmised \cite{Guanella2007b}\nocite{Mathis2012c}-\cite{Wei2013}.
In 3D, our theory makes the experimentally testable prediction that grid cells will have firing fields positioned on a face-centered-cubic lattice or its equally dense non-lattice variant -- a hexagonal close packing.

Unimodal tuning curves with a single preferred stimulus, which are characteristic for place cells or orientation-selective neurons in visual cortex, 
have been extensively studied~\cite{Paradiso1988}\nocite{Seung1993,Zhang1999,Pouget1999,Bethge2002,Wilke2002}-\cite{Brown2006}.
This is also true for multinomial tuning curves that are periodic along orthogonal stimulus axes and generate repeating quadratic (or rectangular) 
activation patterns~\cite{Montemurro2006}\nocite{Fiete2008}-\cite{Mathis2012b}. 
Our results extend these studies by taking general stimulus symmetries into account and provide a powerful link to cryptography and coding 
theory~\cite{Shannon1948}\nocite{Conway1992}-\cite{Gray1998}. 
This leads us to hypothesize that optimal lattices  not only underlie the neural representation of 
physical space, but will also be found in the representation of other high-dimensional sensory or cognitive spaces.

\section*{Model}

\subsection*{Population coding for space} 
We consider the $D$-dimensional space $\mathbb{R}^D$ in which spatial location is denoted by coordinates $x = (x_1, \ldots, x_D) \in \mathbb{R}^D$. 
The animal's position in this space is encoded by $N$ noisy, statistically independent neurons. We account for trial-to-trial variability in neuronal 
firing by assigning a probability $P_i(k_i | \tau \,  \Omega_i(x))$ for neuron $i$ to fire $k_i$ spikes within a fixed time window $\tau$ when the animal is at position $x$. 
$\Omega_i(x)$ describes the mean firing rate of neuron $i$ as a function of $x$ and is called the neuron's ``tuning curve''. 
The conditional probability of the $N$ neurons to fire $(k_1,\ldots,k_N)$ spikes at position $x$ completes the encoding model: \begin{equation}
 P((k_1,\ldots,k_N) | x) = \prod_{i=1}^{N} P_i(k_i | \tau \,  \Omega_i(x)). \label{ProbModel}
\end{equation}
Decoding relies on inverting this conditional probability. 

\begin{figure*}
\includegraphics[angle=0,width=.9\textwidth]{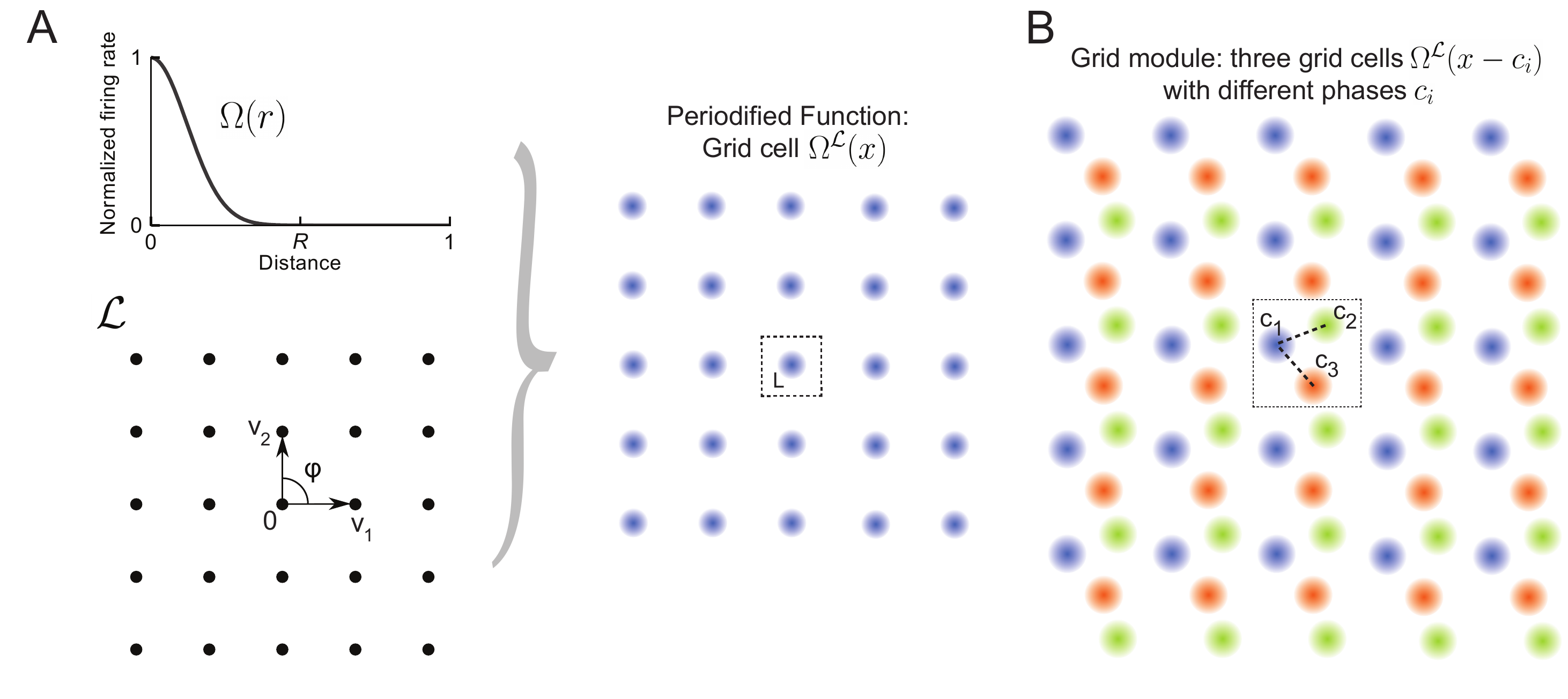}
\caption{Grid cells and modules. {\bf (A)} Construction of a grid cell: Given a tuning shape $\Omega$ and a lattice $\mathcal{L}$, here a square lattice 
generated by $v_1$ and $v_2$ with $\varphi=\pi/2$, one periodifies $\Omega$ 
with respect to $\mathcal{L}$. 
One defines the value of 
$\Omega^\mathcal{L}(x)$ in the fundamental domain $L$ as the value of 
$\Omega(r)$ applied to the distance from zero and then repeats this map over 
$\mathbb{R}^2$ like $\mathcal{L}$ tiles the space. This construction can be 
used for lattices $\mathcal{L}$ of arbitrary dimensions (Eq.~\eqref{eq:periodic_extension}).
{\bf (B)} Grid module: The firing rates of three grid cells (orange, green and blue) are indicated by color intensity. The cells' tuning is 
identical ($\Omega$ and $\mathcal{L}$ are the same). Yet they differ in their spatial phases $c_i$. 
Together, such identically tuned cells with different spatial phases define a grid module.}
\label{fig:Figure1}
\end{figure*}

\subsection*{Periodic tuning curves} Grid cells have periodic tuning curves. The periodic structure of $\Omega_i(x)$ reflects the symmetries of 
the tuning curve, i.e., the set of vectors that map the tuning curve onto itself. We want to understand how the periodic structure 
affects the resolution of the population code. To compare grid cells 
with different periodic structures, we construct periodic  tuning curves as illustrated in Fig.~\!\ref{fig:Figure1}A. 
We start with a lattice $\mathcal{L}$ and a tuning shape $\Omega: \mathbb{R}^+ \rightarrow [0,1]$ that decays from unity to zero; $\Omega$ should at least be twice 
continuously differentiable. Specifically, let $\mathcal{L} \subset \mathbb{R}^D$ be a non-degenerate point lattice,

\begin{equation} \mathcal{L} = \sum_{\alpha = 1}^D k_\alpha v_\alpha
\qquad 
\text{for}  \qquad 
k_\alpha \in \mathbb{Z} ,\; v_\alpha \in \mathbb{R}^D,
\label{eq:lattice_definition}
 \end{equation}
such that $\left(v_\alpha \right)_{1 \leq \alpha \leq D}$ is a basis of $\mathbb{R}^D$. Each lattice point $p \in \mathcal{L}$ has a 
domain $V_{p} \subset \mathbb{R}^D$ (also called Voronoi region), defined as
\begin{equation}
 V_{p} = \{ x \in \mathbb{R}^D \ | \ \|x-p\| < \| x-q\| \ \forall q \in \mathcal{L} \ \wedge \ p\neq q \}. \label{Voronoidefinition}
\end{equation}
The term $\|\ \! \! . \  \!\! \|$ denotes Euclidean distance, $\|x\|=\sqrt{\sum_\alpha x_\alpha^2}$. Note that $V_p \cap V_q = \varnothing$ if $p \neq q$ and 
that for all $p,q \in \mathcal{L}$ there exists a unique vector $v\in \mathcal{L}$ with $V_p = V_q+v$.

The domain that contains the null ($0$) vector is called fundamental domain and denoted by $L:=V_0$. For each $x\in \mathbb{R}^D$ there is a unique
$p\in \mathcal{L}$ such that $x-p \in L$. Let us call this mapping $\pi_{\mathcal{L}}$. Then one can periodify $\Omega$ onto $\mathcal{L}$ by defining 
a grid cell's tuning curve as $\Omega^\mathcal{L}$:
\begin{equation}
\Omega^\mathcal{L} (x) : \mathbb{R}^D \rightarrow \mathbb{R}^+, \ x \mapsto
f_{max} \cdot \Omega(\| \pi_{\mathcal{L}}(x) \|^2), \label{eq:periodic_extension}
\end{equation} where $f_{max}$ is the peak firing rate of the neuron. Note that throughout the paper we set $f_{max}$=$\tau$=1. Within $L$, the 
tuning curve defined above is radially symmetric. This pattern is repeated along the nodes of $\mathcal{L}$, akin to a ceramic tiling.

A grid module is an ensemble of $M$ grid cells $\Omega^\mathcal{L}_i$, $i \in \{ 1, \ldots   M\}$ with identical, 
but spatially shifted tuning curves, i.e., $\Omega^\mathcal{L}_i(x) = \Omega^{\mathcal{L}+c_i}(x)$ and spatial phases $c_i\in L$ (see Fig.~\!\ref{fig:Figure1}B). 
The various phases within a module can be summarized by their phase density $\rho(c) = \sum_i^M \delta(c-c_i)$.  
A module is uniquely characterized by its signature $(\Omega,\rho,\mathcal{L})$. For a given function $\Omega$ and
density $\rho$ we can ask which lattice $\mathcal{L}$ yields the highest resolution. 
To answer this question, we first need to define the resolution of a module in the context of population coding.

\subsection*{Resolution and Fisher information}

Given a response  of $K=\left(k_1,\ldots,k_n\right)$ spikes across the population, we ask how accurately an ideal observer can deduce the stimulus $x$. 
The inverse of the Fisher information (FI) matrix $\bm J(x)$,
\begin{equation}
 J_{\ab}(x) = \int \!\!  \left(\frac{\partial  \ln P(K,x)}{\partial
x_\alpha}
\right)
\left(\frac{\partial \ln P(K,x)}{\partial x_\beta}\right) P(K,x) \mathrm{d}K,
\label{DefFI}
\end{equation}
 bounds  the covariance matrix $\bm \Sigma$ of the estimated coordinates $x = (x_1, \ldots, x_D)$, and thus the resolution any unbiased
estimator of the encoded stimulus can achieve,
\begin{equation}
 \bm \Sigma(x) \geq \bm J(x)^{-1}.
 \label{eq:cramer-rao}
\end{equation}
This is known as Cram\'er-Rao bound~\cite{Lehmann1998}. A homogeneous spatial representation requires that
$\bm J(x)$ be asymptotically independent of $x$ (as $M$ becomes large); spatial isotropy implies  that the diagonal entries in $\bm J(x)$ are equal. 
These two conditions assure that the population 
has the same resolution at any location and gives the same resolution along each axis.

\section*{Results}

To determine how the resolution of a grid module depends on the periodic 
structure $\mathcal{L}$, we compute the population
Fisher information $\bm J_{\varsigma}(x)$ for a module of grid cells with 
signature $\varsigma=(\Omega,\rho,\mathcal{L})$. 
By fixing the tuning shape $\Omega$ and the number $|\rho|=M$ of spatial phases, we can 
calculate the resolution.


\subsection*{Scaling of lattices and nested grid codes}

Our grid-cell construction has one obvious degree of freedom, the length
scale of the lattice $\mathcal{L}$. For a module with signature $\varsigma=(\Omega,\rho,\mathcal{L})$ and for arbitrary scaling factor $\lambda > 0$, $\lambda \varsigma:=(\Omega(\lambda r),\rho(\lambda x),\lambda \cdot \mathcal{L})$ is a grid module, too. 
The corresponding tuning curve satisfies ${(\Omega\circ\lambda)}_{\lambda 
\mathcal{L}}(x) = \Omega_\mathcal{L}(\lambda x)$ and is thus merely a 
scaled version of the former. 
Indeed, as we show in the methods, the FI of the rescaled module is $\lambda^{-2}  \bm J_{\varsigma} (0)$. 

The Cram\'er-Rao bound (Eq.\eqref{eq:cramer-rao}) implies that the resolution of an unbiased estimator could
thus rapidly improve with decreasing $\lambda$. However, the multiple firing fields of a grid cell cannot be distinguished by a decoder, so 
that for $\lambda\to 0$ the global resolution approaches the a-priori uncertainty~\cite{Mathis2012b,Mathis2012a}.
By combining multiple grid modules with different spatial periods one can overcome this fundamental limitation, counteracting the ambiguity caused by periodicity and still preserving 
 the highest resolution at the smallest scale. 
Thus, one arrives at nested populations of grid modules, whose spatial periods 
range from coarse to fine. 
The FI for an individual module at one scale determines the optimal length scale of the next module~\cite{Mathis2012b,Mathis2012a}. 
The larger the FI per module,  
the greater the refinement  at subsequent scales can be~\cite{Mathis2012b,Mathis2012a}.  This fact underscores the importance of  finding the lattice that endows a grid 
module with maximal FI.

This result emphasizes the importance of finding the lattice that endows a grid module with maximal FI, 
but also highlights that the specific scale of the lattices can be fixed for this study. Without loss of generality, we therefore only consider lattices whose inter-nodal distances 
are at least one. 

\begin{figure*}
\centering
\includegraphics[angle=0,width=.8\textwidth]{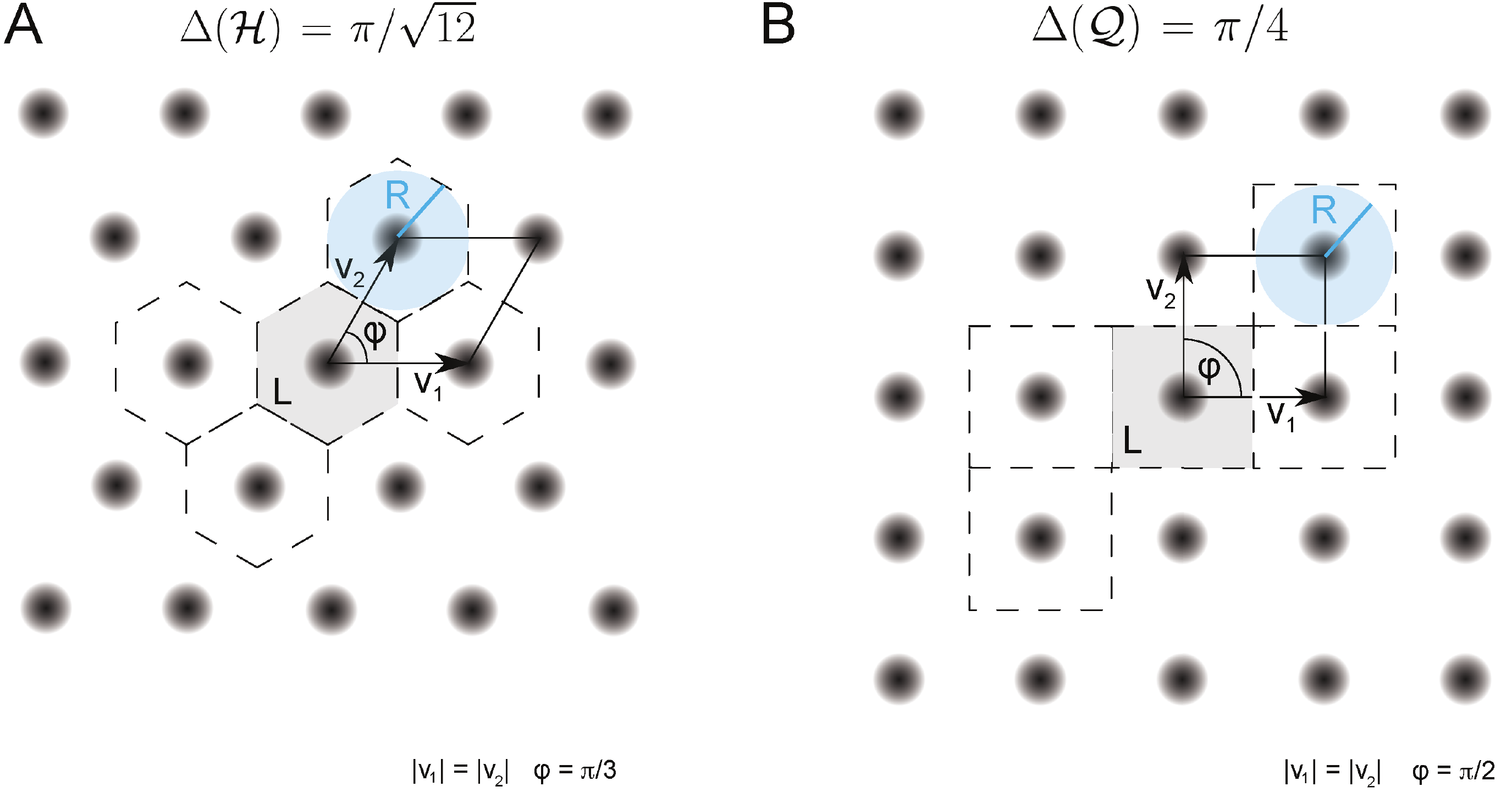}
\caption{Periodified grid-cell tuning curve $\Omega^\mathcal{L}$ for two planar lattices, {\bf (A)} the hexagonal (equilateral triangle)  lattice $\mathcal{H}$ and {\bf (B)} the square lattice $\mathcal{Q}$, together with the basis vectors $v_1$ and $v_2$. 
These are $\pi/3$ 
apart for the hexagonal lattice and $\pi/2$ for the square lattice. The fundamental domain, i.e., the Voronoi cell around $0$,
is shown in gray. 
A few other domains that have been generated according to the lattice symmetries are marked by dashed lines. The blue disk shows the disk with maximal radius $R$ that can be inscribed in the two fundamental domains. 
For equal and unitary node-to-node distances, i.e. $|v_1|=|v_2| = 1$, the maximal radius equals $1/2$ for both lattices. 
The packing ratio $\Delta$ 
is  $\Delta(\mathcal{H}) = \pi/\sqrt{12}$ for the hexagonal and $\Delta(\mathcal{Q}) = \pi/4$ for the square lattice; the hexagonal lattice is approximately $15.5\%$ denser than the square lattice.}
\label{fig:Figure2}
\end{figure*}

\subsection*{Fisher information of a grid module with lattice $\mathcal{L}$}

If the grid-cell density $\rho$ is uniform across $\mathcal{L}$, 
then for all $x\in \mathbb{R}^D$: $\bm
J_{\varsigma}(x) \approx \bm J_{\varsigma}(0)$. It therefore suffices to only consider 
the FI at the origin.
 Furthermore, for  cells whose firing is statistically independent (compare Eq.~\eqref{ProbModel}), the joint 
probability factorizes;
thus, the population FI is just the sum over the individual 
neuron contributions,

\begin{equation}
 \bm J_{\varsigma}(0) = \sum_i \bm J_{\Omega_i^\mathcal{L}}(0) =  \sum_i
\bm J_{\Omega^\mathcal{L}}(c_i) = \int_L \bm J_{\Omega^\mathcal{L}}(c) \rho(c)
\mathrm{d}c. \label{PopulationFisherinformationDef}\end{equation}

For increasing size $M$ of a module with uniformly distributed neurons the law of large numbers implies
\begin{equation} \lim_{M \rightarrow \infty } \left| \frac{\mathrm{det}(\mathcal{L})}{M} \bm
J_{\varsigma}(0) - \int_L  \bm J_{\Omega^\mathcal{L}}(c)  \mathrm{d}c
\right| = 0.     \end{equation}
Here, $\mathrm{det}(\mathcal{L})$ denotes the volume of the fundamental domain.
Thus for large numbers of neurons $M=\int_L \rho(c) \mathrm{d}c$ we obtain

\begin{equation}
 \bm J_{\varsigma}(0) \approx \frac{M}{\mathrm{det}(\mathcal{L})} \int_L 
\bm J_{\Omega^\mathcal{L}}(c) \mathrm{d}c.
\end{equation}

Let us assume that $\mathrm{supp} (\Omega) = [0,R]$ for some positive radius $R$.
Grid cells that have a non-vanishing FI contribution
to $x=0$  are thus contained in the $R$-ball $B_R(0)$. 
If we now also assume that
$B_R(0) \subset L $, we get 

\begin{equation}
 \int_{L} \bm J_{\Omega^\mathcal{L}}(c) \mathrm{d}c = \int_{B_R(0)}
\bm J_{\Omega^\mathcal{L}}(c) \mathrm{d}c.
\end{equation}
This result implies that any grid code $\varsigma=(\Omega,\rho,\mathcal{L})$, 
with $M\gg 0$, $\mathrm{ supp} (\Omega) = [0,R]$ 
and $B_R(0) \subset L $, satisfies

\begin{equation}
\bm J_{\varsigma}(0) \approx \frac{M}{\mathrm{det}(\mathcal{L})}
\int_{B_R(0)}
\bm J_{\Omega^\mathcal{L}}(c) \mathrm{d}c.
 \label{PopulationFisherinformation}
\end{equation}
The FI at the origin is thus approximately equal to the product of the mean 
FI contribution of cells within a $R$-ball around $0$ and the 
number of neurons $M$, weighted by the ratio of the volume of the $R$-ball 
to the area of the fundamental domain $L$.  By the radial symmetry of $\Omega^\mathcal{L}$,  $J_{\Omega^\mathcal{L}}(c)$ is diagonal with identical entries, guaranteeing 
 the spatial resolution's isotropy. 

For two lattices $\mathcal{L}_1$,$\mathcal{L}_2$, with $B_R(0) 
\subset L_1 \cap L_2$ we consequently obtain

\begin{equation}
\frac{\mathrm{tr} \bm J_{\Omega^{\mathcal{L}_1}}}{\mathrm{tr} \bm
J_{\Omega^{\mathcal{L}_2}}} = \frac{
\mathrm{det}(\mathcal{L}_2)}{\mathrm{det}(\mathcal{L}_1)}, \label{inverselyprop}
\end{equation}
which means that the resolution of the grid codes is inversely proportional to 
the volumes of their fundamental domains. 
This result implies that finding the maximum FI translates directly  into finding the lattice with the highest packing ratio. 

\subsection*{Packing ratio of lattices}

The sphere packing problem is of general interest in mathematics~\cite{Conway1992} and  has wide-ranging 
applications from crystallography to information theory~\cite{Shannon1948,Gray1998,Barlow1883,Gruber2003}. 
When packing $R$-balls $B_R$ in $\mathbb{R}^D$ in an non-overlapping fashion, the density of the packing is defined 
as the fraction of the space covered by balls. For a lattice $\mathcal{L}$, it is given by

\begin{equation}
\frac{\mathrm{vol}(B_R(0))}{\mathrm{det}(\mathcal{L})}, \label{eq:defpackingratio}\end{equation} 
which is known as the packing ratio $\Delta(\mathcal{L})$ of the lattice. 
For a given lattice, this ratio is maximized by choosing the largest possible $R$, 
known as the packing radius, which is defined as the in-radius of a Voronoi region containing the 
origin~\cite{Conway1992}. 

\subsection*{Fisher information and packing ratio}

We now come to the main finding of this study: Among grid modules with different lattices,
the lattice with the highest packing ratio leads to the highest spatial resolution.

\begin{figure*}
\centering
\includegraphics[angle=0,width=.9\textwidth]{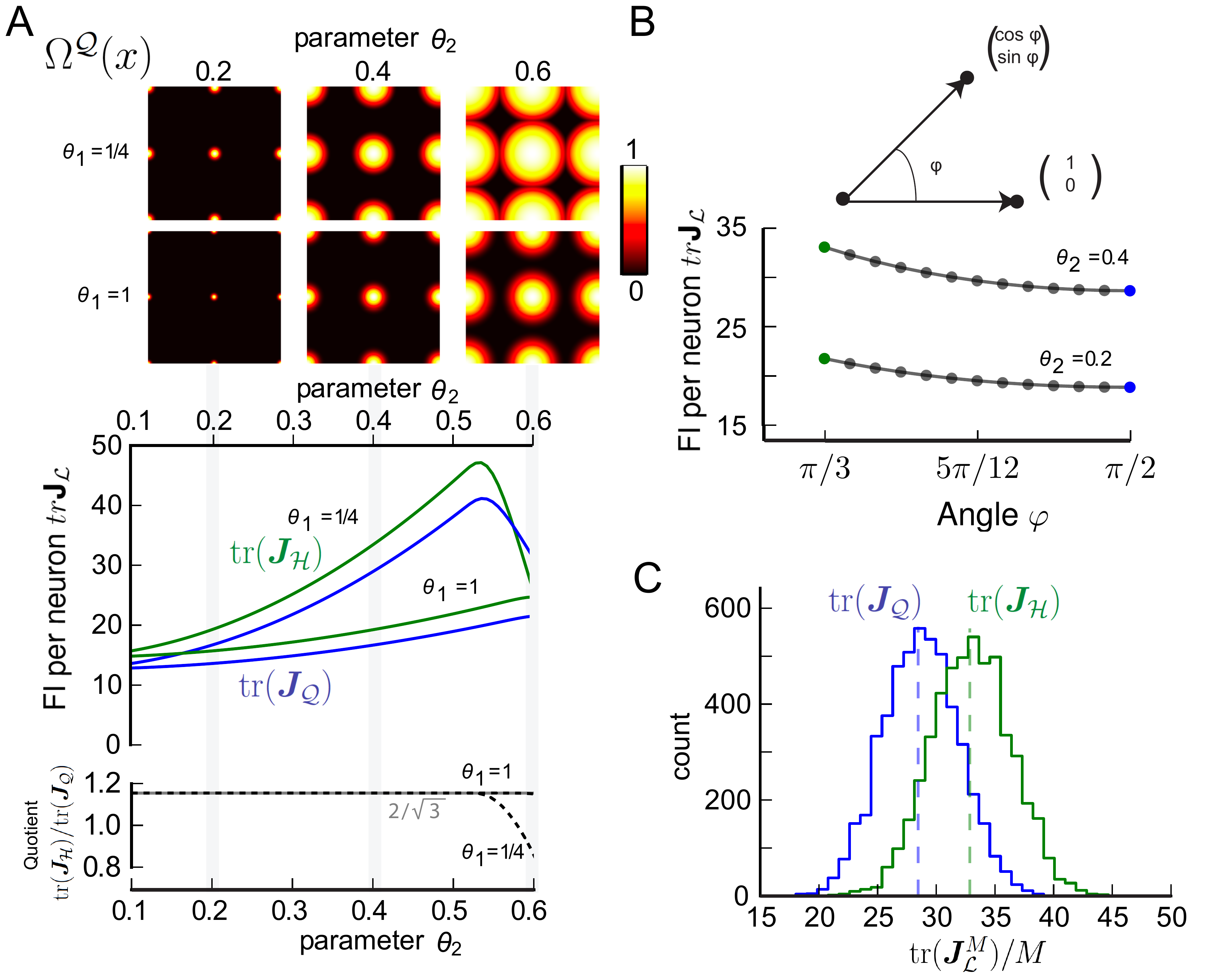}
\caption{Fisher information for modules of 2D grid cells. {\bf (A) Top:} Periodified bump-function $\Omega$ and
square lattice $\mathcal{L}$, for various parameter combinations $\theta_1$ and $\theta_2$.
Here, $\theta_1$ modulates the decay and $\theta_2$ the support. 
{\bf Middle:} Average trace $\textrm{tr} \bm J_{\mathcal{L}}$ of FI for uniformly distributed grid cells $\Omega^\mathcal{L}$.
Hexagonal ($\mathcal{H}$) and square ($\mathcal{Q}$) lattices are considered for different $\theta_1$ and $\theta_2$ values. 
The FI of the hexagonal grid cells outperforms the quadratic grid when support is fully within the fundamental domain ($\theta_2<0.5$, see main text). 
{\bf Bottom:} Ratio $\textrm{tr} \bm J_{\mathcal{H}} / \textrm{tr} \bm J_{\mathcal{Q}}$ as function of the tuning parameter $\theta_2$. 
For $\theta_2<0.5$, the hexagonal population offers $\sqrt{3}/2$ times the resolution of the square population, as predicted 
by the respective packing ratios. 
{\bf (B)} Average $\textrm{tr} \bm J_{\mathcal{L}}$ for grid cells distributed uniformly in lattices generated by basis vectors separated by an angle $\varphi$ (basis depicted above graph). 
$\textrm{tr} \bm J_{\mathcal{L}}$ behaves like $1/\sin(\varphi)$ and has its maximum at $\pi/3$. 
{\bf (C)} Distribution of $5,000$ realizations of $\textrm{tr} \bm J^M_{\mathcal{L}}/M$ at $0$ for a 
population of $M=200$ randomly distributed neurons.
For both the hexagonal and square lattice, parameters are $\theta_1=1/4$ and $\theta_2=0.4$. 
The means closely match the average values in (A). 
However, due to the finite neuron number the FI varies strongly for different realizations, and in about $20\%$ of the cases a square lattice module outperforms a hexagonal lattice.}
\label{fig:Results2D}
\end{figure*}

To derive this result, let us fix a tuning 
shape
$\Omega$ with $\mathrm{ supp} 
(\Omega) = [0,R]$, lattices $\mathcal{L}_j$ such that $B_R(0) \subset L_j$ for 
$1 \!\leq\! j \!\leq\! K$ and uniform densities $\rho$ for each fundamental domain of 
equal cardinality $M$. 
Any linear order on the packing ratios,
\begin{equation}
 \Delta(\mathcal{L}_1) \leq \ldots \leq \Delta(\mathcal{L}_j) \leq \ldots \leq
\Delta(\mathcal{L}_K) \ ,
\end{equation}
is translated by Eq.~\eqref{inverselyprop} into the same order for the traces 
of the FI
\begin{equation}
 \mathrm{tr} \bm J_{\Omega^{\mathcal{L}_1}} \leq \ldots \leq \mathrm{tr} \bm
J_{\Omega^{\mathcal{L}_j}} \leq \ldots \leq \mathrm{tr} \bm
J_{\Omega^{\mathcal{L}_K}} \label{mainresult}
\end{equation}
and thus the resolution of these modules: the higher the 
packing ratio, the higher the FI of a grid module. 

Before moving on to the implications of this finding, let us explain why the condition $\mathrm{ supp} (\Omega) = [0,R]$ with $B_R(0) \subset L$,  although restrictive, 
is biologically plausible. Rodent experiments show that grid cells tend to stop firing between grid fields and that the typical ratio between field width and spatial period is 
well below $0.5$~\cite{Hafting2005,Brun2008,Giocomo2011b}. 

The optimal packing ratio 
in various spaces is well known. 
Having established our main result, we can now draw on a rich body of literature, in particular~\cite{Conway1992}, to discuss 
the expected firing-field structure of grid cells in 2D and 3D environments.

\subsection*{Optimal two-dimensional grid cells}

With a packing ratio of $\pi / \sqrt{12}$, the hexagonal lattice is the densest lattice in the plane~\cite{Lagrange1773}.
Accordingly, Eq.~\eqref{inverselyprop}
implies that the hexagonal lattice is the optimal arrangement for grid-cell firing fields on the plane.
It outperforms the quadratic lattice (density $\pi / 4$) by about $15.5\%$ 
(see Fig.~\!\ref{fig:Figure2}). 
Consequently, the FI of a grid module periodified along a hexagonal lattice 
outperforms one periodified along a square lattice by the same factor.

For concreteness, we calculated the trace of the average Fisher 
information $\textrm{tr} \bm J_{\varsigma}/\int_L \rho$ for signature $\varsigma = 
(\Omega,\rho,\mathcal{L})$ and chose the lattice $\mathcal{L}$ to either be the 
hexagonal lattice $\mathcal{H}$ or the quadratic lattice $\mathcal{Q}$.
We denote the trace of the average FI per neuron as: 
$\textrm{tr} \bm J_{\mathcal{L}}$ = $\textrm{tr} \bm J_{\varsigma}/\int_L \rho$; $\textrm{tr} \bm J_{\mathcal{H}}$ and $\textrm{tr} \bm J_{\mathcal{Q}}$ are 
similarly defined. 
We considered Poisson spike statistics
and used a bump-like tuning shape $\Omega$ (Eq.~\eqref{TuningcurveBump} in the methods section).
$\Omega$ depends on two parameters $\theta_1$ and $\theta_2$, 
where $\theta_1$ controls the slope of the flank in $\Omega$ and $\theta_2$ defines the support radius.  
The periodified tuning curve $\Omega^\mathcal{Q}$ is shown for different parameters in the top of Fig.~\ref{fig:Results2D}A 
and Fig.~\!\ref{fig:IllustrationBump} in the methods. 

Fig.~\ref{fig:Results2D}A shows $\textrm{tr} \bm 
J_{\mathcal{H}}$ and $\textrm{tr} \bm J_{\mathcal{Q}}$ 
for various values of $\theta_1$ and $\theta_2$. 
Quite generally, the FI is larger for grid modules with broad tuning (large $\theta_2$) and 
steep tuning slopes (small $\theta_1$).
Fig.~\ref{fig:Results2D}A also demonstrates that as long as $\theta_2\leq 1/2$, 
$\textrm{tr} \bm J_{\mathcal{H}}$ consistently outperforms $\textrm{tr} \bm J_{\mathcal{Q}}$. But how large is this effect?
As predicted by the theory,
the grid module with the hexagonal lattice outperforms the square lattice by the relation of 
packing ratios $\sqrt{3}/2$, as long as the support radius $\theta_2$ is within 
the fundamental domain of the hexagonal and the square lattice of unit length, 
i.e. $\theta_2\leq 1/2$ (bottom of Fig.~\ref{fig:Results2D}A shows that). 
As the support radius becomes larger, the FI of the hexagonal lattice is no longer necessarily greater than that of the  square lattice.
For tuning curves with larger support boundary effects influence the FI and which lattice is better depends on the specific interplay of tuning curve and boundary shape: 
for $\theta_1=1/4$, $\textrm{tr} \bm J_{\mathcal{H}}/\textrm{tr} \bm J_{\mathcal{Q}}$ drops quickly  beyond $\theta_2=0.5$, 
even though, for $\theta_1 = 1$, the ratio  stays constant up to $\theta_2=0.6$.

Next we calculated FI per neuron for a larger family of planar lattices generated by two unitary basis vectors with angle $\varphi$. 
Fig.~\!\ref{fig:Results2D}B displays $\textrm{tr} \bm J_{\mathcal{L}}$ for $\varphi \in [\pi/3,\pi/2]$,
slope parameter $\theta_1=1/4$ and different support radii $\theta_2$. 
The value $\varphi=\pi/3$ is the lower limit for the lattice to have unitary length. 
The $\textrm{tr} \bm J_{\mathcal{L}}$ decays with increasing angle $\varphi$. 
Indeed, as suggested by Eq.~\eqref{PopulationFisherinformation} the FI falls like $1/\det{\mathcal{L}} = 1/\sin(\varphi)$ so that the maximum is achieved for the hexagonal lattice with $\pi/3$.

The FIs $\textrm{tr} \bm J_{\mathcal{L}}$ are averages over all phases, under the assumption that the density of phases tends to a constant; 
are these values also indicative for small neural populations? 
To answer this question, we calculated the FI for populations with $200$ neurons, as 
one class of putative grid cells is arranged in patches of of this size~\cite{Ray2014}. 
For $M=200$ randomly chosen phases (Fig.~\!\ref{fig:Results2D}C), the mean of the normalized Fisher Informations $\textrm{tr} \bm J_{\mathcal{L}}^M / M$ over $5,000$ 
realizations is well captured by the FI per neuron calculated in Fig.~\!\ref{fig:Results2D}A. Because of fluctuations in the FI, however, 
the square lattice is better than the hexagonal lattice in about $20\%$ of the cases.

Our theory implies that for radially symmetric tuning curves the hexagonal lattice provides the best resolution among all planar lattices. 
This generalizes results of other authors who considered a notion of resolution defined as the range of the population code per smallest distinguishable scale~\cite{Wei2013} 
or compared hexagonal and quadratic lattices based on numerical maximum likelihood 
decoding~\cite{Guanella2007b}.

\subsection*{Optimal three-dimensional grid cells} 

\begin{figure*}
\centering
\includegraphics[angle=0,width=.9\textwidth]{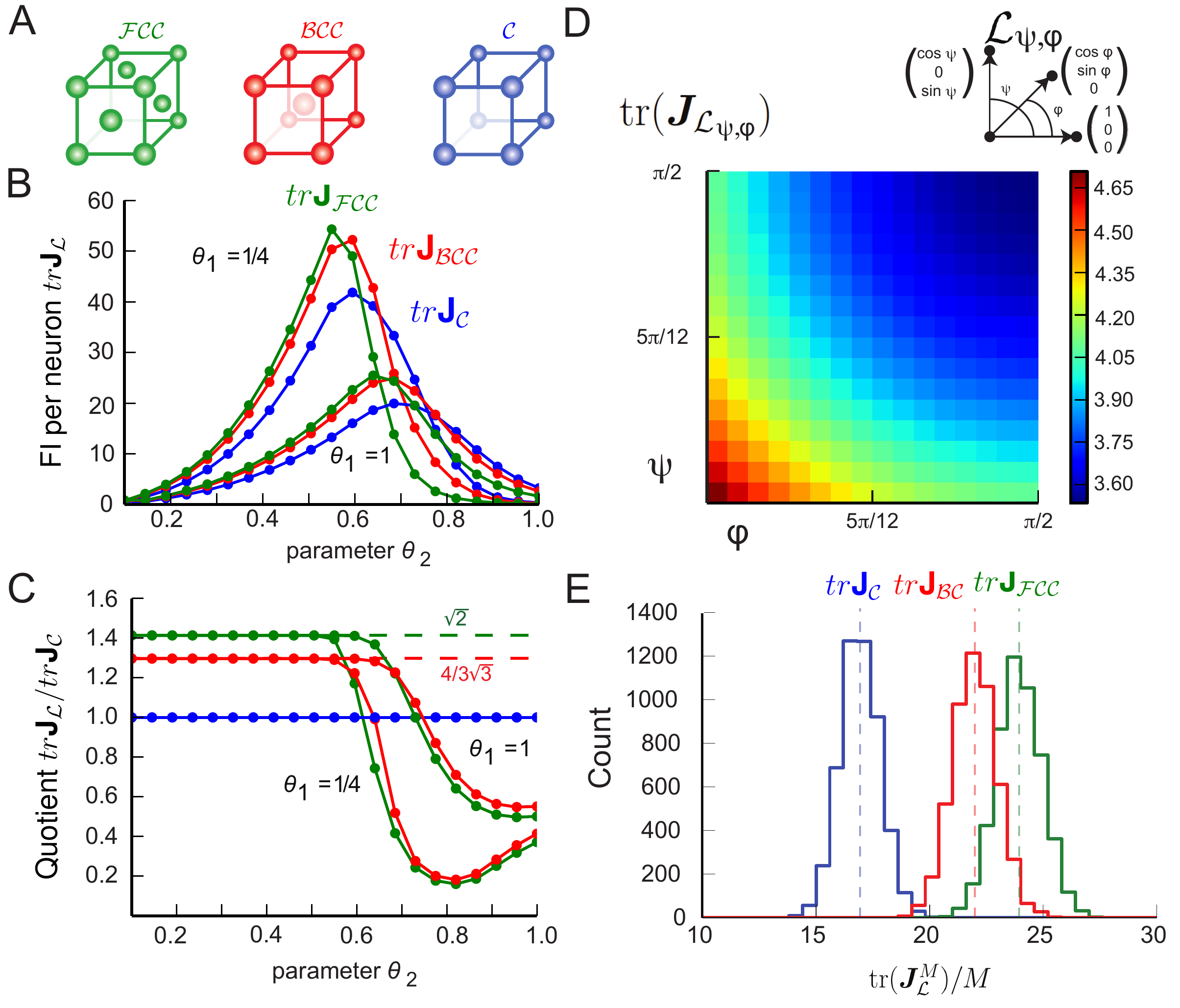}
\caption{Fisher information for modules of 3D grid cells. {\bf (A)} The three  lattices considered: face-centered cubic ($\mathcal{FCC}$),
body-centered cubic ($\mathcal{BCC}$) and cubic ($\mathcal{C}$). 
{\bf (B)} $\textrm{tr} \bm J_{\mathcal{L}}$ for the periodified bump-function $\Omega$ for the three lattices and various parameter combinations $\theta_1$ and $\theta_2$. 
The FI of the $\mathcal{FCC}$ grid cells outperforms the other lattices when the support is fully within the fundamental domain ($\theta_2<0.5$, see main text). For larger $\theta_2$ the best lattice depends on the relation between the Voronoi cell's boundary and the tuning curve.
{\bf (C)} Ratio $\textrm{tr} \bm J_{\mathcal{L}} / \textrm{tr} \bm J_{\mathcal{C}}$ as a function of $\theta_2$ for $\mathcal{L} \in \{ \mathcal{FCC}, \mathcal{BCC}, \mathcal{C}\}$. 
For $\theta_2<0.5$, the hexagonal population has $\sqrt{3}/2$ times the resolution of the square population, as predicted by the packing ratios. 
{\bf (D)} Average $\textrm{tr} \bm J_{\mathcal{L_{\varphi,\psi}}}$ for uniformly distributed grid cells within a lattice $\mathcal{L}_{\varphi,\psi}$ generated by basis vectors separated by angles $\varphi$ and $\psi$ (as shown above; $\theta_1=\theta_2=1/4$). 
$\textrm{tr} \bm J_{\mathcal{L}_{\varphi,\psi}}$ behaves like $1/(\sin \varphi \cdot\sin \psi)$ 
and has its maximum for the lattice with the smallest volume. 
{\bf (E)} Distribution of $5,000$ realizations of $\textrm{tr} \bm J^M_{\mathcal{L}}/M$ at $0$ for a population of $M=200$ randomly distributed neurons. 
Parameters: $\theta_1=1/4$, $\theta_2=0.4$. 
The means closely match the averages in (B). Due to the finite neuron number the FI varies strongly for different realizations. }
\label{fig:Results3D}
\end{figure*}

Gauss proved that the packing ratio of any cubic lattice is bounded by $\pi / (3\sqrt{2})$ and that this value is attained for the face-centered
cubic ($\mathcal{FCC}$) lattice~\cite{Gauss1826} illustrated in Fig.~\!\ref{fig:Results3D}A. 
This implies that the optimal 3D grid-cell tuning is given by 
the $\mathcal{FCC}$ lattice. 
For comparison, we also calculated the average population FI for two other 
important 3D lattices: the cubic lattice ($\mathcal{C}$), and the body-centered cubic lattice ($\mathcal{BCC}$), both shown in Fig.~\!\ref{fig:Results3D}A.

Keeping the bump-like tuning shape $\Omega$ and independent Poisson noise,
we compared the resolution of grid modules with such lattices (Fig.~\!\ref{fig:Results3D}B). 
Their averaged trace of FI is denoted by $\textrm{tr} \bm J_{\mathcal{FCC}}$, $\textrm{tr} \bm J_{\mathcal{BCC}}$ and $\textrm{tr} \bm J_{\mathcal{C}}$, respectively. 
As long as the support $\theta_2$ of $\Omega$ is smaller than $1/2$, the support is a subset of the fundamental domain of all three lattices. Hence,
the trace of the population FI of the $\mathcal{FCC}$ outperforms both the $\mathcal{BCC}$ and 
$\mathcal{C}$ lattices. As the ratios of
 the trace of the population FI
scales with  the packing ratio (Fig.~\!\ref{fig:Results3D}C), $\mathcal{FCC}$-grid cells provide roughly $40 \%$ more resolution for 
the same number of neurons than cubic lattice grid cells do. Similarly, $\mathcal{FCC}$-grid cells provides $8.8\%$ more 
FI than $\mathcal{BCC}$-grid cells.

Next we calculated the FI per neuron for a large family of cubic lattices $\mathcal{L}_{\varphi,\psi}$ generated by two unitary basis vectors with spanning 
angles $\varphi$ and $\psi$. Fig.~\ref{fig:Results3D}D displays $\textrm{tr} \bm J_{\mathcal{L_{\varphi,\psi}}}$
for $\theta_1=\theta_2=1/4$ and various $\varphi$ and $\psi$. The resolution
$\textrm{tr} \bm J_{\mathcal{L}}$  decays with increasing angles and has its maximum for the lattice with the smallest
volume as predicted by by Eq.~\eqref{PopulationFisherinformation}. 

To study finite-size effects, we  simulated $5,000$  populations of $200$ grid cells with random spatial phases.
Qualitatively, the results (Fig.~\!\ref{fig:Results3D}E) match those in 2D (Fig.~\!\ref{fig:Results2D}C). Despite the small module size, $\mathcal{FCC}$ outperformed 
the cubic lattice $\mathcal{C}$ in all simulated realizations.

\subsection*{Equally optimal non-lattice solutions}

\begin{figure*}
\centering
\includegraphics[angle=0,width=0.9\textwidth]{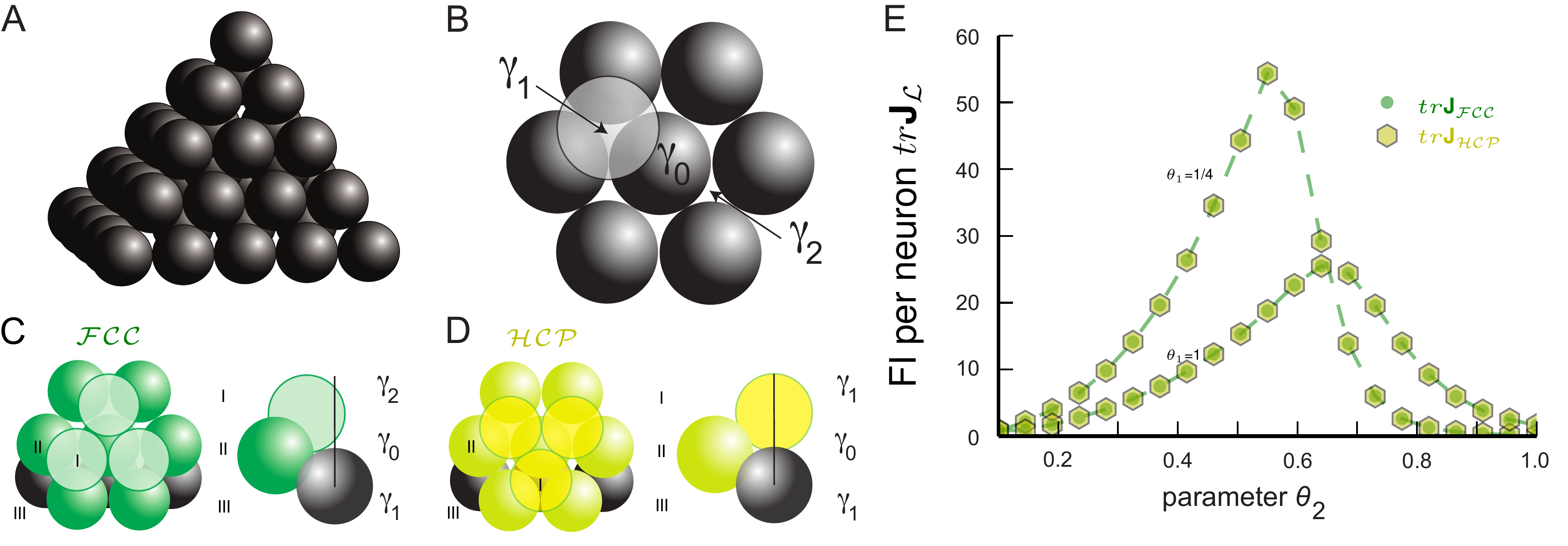}
\caption{Lattice and non-lattice solutions in 3D. {\bf (A)} Stacking of spheres as in a $\mathcal{FCC}$ lattice.
In this densest lattice in 3D, each sphere touches $12$ other spheres and there are four different planar hexagonal lattices through each node.
{\bf (B)} Over a layer of hexagonally arranged spheres centered at $\gamma_0$ (in black) one can put another hexagonal layer by starting from one of six locations, two of which are highlighted, $\gamma_1$ and $\gamma_2$. 
{\bf (C)} If one arranges the hexagonal layers according to the sequence $(\ldots,\gamma_1,\gamma_0,\gamma_2,\ldots)$ one obtains the $\mathcal{FCC}$. 
Note that spheres in layer I are not aligned with those in layer III. 
{\bf (D)}  Arranging the hexagonal layers following the sequence $(\ldots,\gamma_0,\gamma_1,\gamma_0,\ldots)$ leads to the hexagonal close packing 
$\mathcal{HCP}$. 
Again, each sphere touches $12$ other spheres. 
However, there is only one plane through each node for which the arrangement of the centers of the spheres is a regular hexagonal lattice. 
This packing has the same packing ratio as the $\mathcal{FCC}$, but is not a lattice. 
{\bf (E)} $\textrm{tr} \bm J_{\mathcal{L}}$ for bump-function $\Omega$ with $\mathcal{L}=\mathcal{FCC}$ and $\mathcal{HCP}$  for various parameter 
combinations $\theta_1$ and $\theta_2$; $\theta_1$ modulates the decay and $\theta_2$ the support. The two packings have the same packing ratio and for this tuning curve also provide 
identical spatial resolution.} 
\label{fig:Results3Dpackings}
\end{figure*}

Fruit is often arranged as $\mathcal{FCC}$ (Fig.~\!\ref{fig:Results3Dpackings}A). 
One arrives at this lattice by starting from a layer of hexagonally placed spheres.
This requires two basis vectors to be specified and is the densest packing in 2D. 
To maximize the packing ratio in 3D a next layer of hexagonally arranged spheres has to be stacked as tightly as possible. 
Modulo hexagonal symmetry,  two choices for the third and final basis vector achieve this packing, denoted as $\gamma_1$ and $\gamma_2$ in Fig.~\!\ref{fig:Results3Dpackings}B. 
If one chooses $\gamma_1$, then two layers below there is no sphere with center at location $\gamma_1$, but instead there is one at
 $\gamma_2$ (and vice versa). This stacking of layers is shown in 
Fig.~\!\ref{fig:Results3Dpackings}C. 

One could  achieve the same density by picking $\gamma_1$ for both the top layer and the layer below the basis layer. 
Yet as this arrangement, called hexagonal close packing (HCP)  cannot be described by three vectors, it does not define a lattice 
(see Fig.~\!\ref{fig:Results3Dpackings}D), even though it is as tightly packed as the $\mathcal{FCC}$. Such packings,  defined as an arrangement of equal non-overlapping balls~\cite{Conway1992,Hales2012}, generalize lattices. For a given packing $\mathcal{P}$ of $\mathbb{R}^D$ by
balls $B_1$ of radius $1$, one can also define a ``grid cell'' by generalizing the definition given for lattices (Eq.~\eqref{eq:periodic_extension}). 
To this end, consider the Voronoi partition of $\mathbb{R}^D$ by $\mathcal{P}$. For each
location $x\in\mathbb{R}^D$ there is a unique Voronoi cell $V_p$ with node $p \in \mathcal{P}$. 
One defines the grid cell's tuning curve $\Omega^\mathcal{P}(x)$ by assigning the firing rate according to $\Omega(\| p- x \|^2 )$ for tuning shape $\Omega$ and
distance $\| p- x \|$. 
This tuning curve $\Omega^{\mathcal{P}}$ is not necessarily periodic, but may satisfy many symmetries, which are determined by the packing $\mathcal{P}$. 
For example, the $\mathcal{HCP}$ defines a highly symmetric packing, which can be used to define a grid cell $\Omega^\mathcal{HCP}(x)$. 
Indeed, one can calculate the trace of the average FI for a module of 
$\mathcal{HCP}$ grid cells and compare it to the $\mathcal{FCC}$ case. 
For bump-like tuning curves $\Omega$, both FIs are identical 
(Fig.~\!\ref{fig:Results3Dpackings}E) as expected from the radial symmetry of $\Omega$. 
As a consequence, grid cells defined by either ${\mathcal{HCP}}$ or ${\mathcal{FCC}}$ symmetries provide optimal resolution. 

Figs.~\!\ref{fig:Results3Dpackings}E and~\ref{fig:Results3Dpackings}D show that the cyclic sequences $(\gamma_0,\gamma_1)$ and $(\gamma_1,\gamma_0,\gamma_2)$ lead to $\mathcal{HCP}$ and $\mathcal{FCC}$, respectively.
The centers $\gamma_0$,$\gamma_1$ and $\gamma_2$ can also be used to make a final point on packings: 
There are infinitely many distinct packings with the same density $\pi/(3\sqrt{2})$. 
They can be constructed by inequivalent words, generated by finite walks through the 
triangle with letters $\gamma_0$,$\gamma_1$ and $\gamma_2$~\cite{Hales2012}. 
For instance, $(\gamma_0,\gamma_1,\gamma_0,\gamma_2)$ describes another packing with the same density. 
All packings share one feature:  around each sphere there are exactly $12$ spheres, arranged either in $\mathcal{HCP}$ or $\mathcal{FCC}$ lattice fashion~\cite{Hales2012}. 
Only recently has it been proven that no other arrangement has a higher packing ratio than the $\mathcal{FCC}$, a problem known as Kepler's conjecture~\cite{Hales2005,Hales2012}. Based on these 
results and our comparison of $\textrm{tr} \bm J_{\mathcal{HCP}}$ and $\textrm{tr} \bm J_{\mathcal{FCC}}$ (Fig.~\!\ref{fig:Results3Dpackings}E) we predict that 
3D grid cells will correspond to one of these packings

While there are equally dense packings in 3D, this is not the case in 2D. Thue proved that 
 the hexagonal lattice is unique in being  the densest  amongst all planar packings~\cite{Thue1910};  so  grid cells in 2D should possess a hexagonal lattice structure. 

\section*{Discussion}

Grid cells are active when an animal is near one of any number of multiple locations 
that correspond to the vertices of a planar hexagonal lattice~\cite{Hafting2005}.
 We generalize the notion of a ``grid cell" to arbitrary dimensions, such that a grid cell's 
 stochastic activity is modulated in a spatially periodic manner within  $\mathbb{R}^D$. 
The periodicity is captured by the symmetry group of the underlying lattice $\mathcal{L}$. 
A ``grid module" consists of multiple  cells with equal spatial period but different spatial phases.
Using information theory, we then asked which lattice offers the highest spatial resolution. 

We find that the resolution of a grid module is related to the packing ratio of $\mathcal{L}$ 
-- the lattice with highest packing ratio corresponds to the grid module with highest resolution. 
Well-known results from mathematics~\cite{Conway1992,Lagrange1773,Gauss1826} then show that the hexagonal lattice is optimal for 
representing 2D, whereas the face-centered-cubic ($\mathcal{FCC}$) lattice is optimal for 3D. 
 In 3D, but not in  2D, there are also non-lattice packings with the same resolution
as the densest lattice~\cite{Hales2012,Thue1910}. 
A common feature of these highly symmetric optimal solutions in 3D is that each grid field is surrounded  by $12$ other grid fields, arranged in  
either $\mathcal{FCC}$ lattice or hexagonal close packing fashion. 
These solutions  emerge from the set of all possible packings simply by maximizing the resolution, as we showed.
However, resolution alone does not distinguish between
optimal packing solutions with different symmetries. Whether a realistic neuronal decoder, such as one based on  population vector averages, favors one 
particular solution is an interesting open question.

Maximizing the resolution explains the observed hexagonal patterns of grid cells in two dimensions, and predicts an $\mathcal{FCC}$ lattice 
(or equivalent packing) for grid-cell tuning curves of mammals that can freely explore the three-dimensional nature of their environment.
Quantitatively, we demonstrated that these optimal populations provide $15.5\%$ (2D) and about $41\%$ (3D) more resolution than grid codes with quadratic or cubic grid cells for 
the same number of neurons. Although better, this might not seem substantial,  at least not at the level of a single grid module.  However,  for a nested grid code of $10$ modules, based on estimates 
from the medial entorhinal cortex~\cite{Stensola2012}, this could translate into a gain of $1.155^{10} \approx 4.2$ and 
$\sqrt 2^{10} = 32$, 
respectively~\cite{Mathis2012b,Mathis2012a}.

In this study, we focused on optimizing grid modules for an isotropic and homogeneous space, which means that the resolution should be equal everywhere and in each direction of space. 
From a mathematical point of view, this is the most general setting, but it is certainly not the only imaginable scenario; future studies should shed light on other geometries.
Indeed, the topology of natural habitats, such as burrows or caves, can be highly complicated.  Higher resolution might be required  at spatial locations of  behavioral relevance.  Neural representations of 3D space may also be composed of multiple 1D and 2D patches~\cite{Jeffery2013}. However, the mere fact that these habitats involve complicated low-dimensional geometries does not imply that an animal cannot acquire a general map for the environment.
 Poincar\'e already suggested that an isotropic and homogeneous representation for space can emerge out of non-Euclidean perceptual spaces, as one can move through physical space by 
learning the motion group~\cite{Poincare1913}. However, from an efficient-coding view alone~\cite{Barlow1959,Atick1992,Simoncelli2001}, there is no obvious reason why 
animals should acquire this full representation of space. Experimental evidence suggests that rats do not encode 3D space in an isotropic manner~\cite{Hayman2011}, which could be a  
consequence of the anisotropic way rats had to navigate in this study 
(peg board and helical track), rather than their general conception of space. Data from flying bats, on the other hand,  demonstrate that,
 at least in this species, place cells represent 3D space in a
uniform and nearly 
isotropic manner~\cite{Yartsev2013}.
Our theoretical analysis assumes that the same is true for bat grid cells  and that they have radially symmetric firing fields. From these assumptions, we showed 
that the best arrangement for the  grid-cells' firing fields would be on a  $\mathcal{FCC}$ lattice or in a $\mathcal{HCP}$ packing. Interestingly,  generalizing a self-organizing 
network model for 2D~\cite{Kropff2008} predicts that the very same solutions evolve dynamically in the 3D system~\cite{Stella2013}.

The  majority of spatially modulated cells in rat medial entorhinal cortex have hexagonal tuning curves, but some have firing fields that are spatially periodic bands~\cite{Krupic2012}.
The orientation of these bands tends to coincide with one of the lattice vectors of the grid cells (as the lattices for different grid cells share a common orientation),
 so band cells might be a layout `defect'.
In this context, we should point out  that the lattice solutions are not globally optimal. 
For instance, in 2D, a higher resolution can result from two systems of nested 1D grid codes, which are 
aligned to the $x$ and $y$ axis, respectively, than from a lattice solution
with the same number of neurons. 
The 1D cells would  behave like band cells. Similar counterexamples can be given in higher dimensions, too. 
Radial symmetry of the tuning curve may also be non-optimal. For example, two sets of elliptically tuned 
2D unimodal cells, with orthogonal short axes, typically outperform unimodal cells with radially symmetric tuning curves~\cite{Wilke2002}. Why experimentally observed place fields and other tuning curves seem to be  isotropically tuned is an open question~\cite{O'KeefeJohnDostrovsky1971,Yartsev2013}. 

More generally, we hypothesize that there are species-specific grid-cell representations of 3D space~\cite{Las2013}.
For example, surface-bound animals might not  encode the third dimension 
to the same degree as flying animals, which could  explain 
the anisotropic tuning for 3D in rodents~\cite{Hayman2011}. Similarly, desert ants represent space only as a projection to flat space~\cite{Wohlgemuth2001,Grah2007}. 
An isotropic and homogeneous representation of 3D-space, on the other hand,  facilitates (mental) rotations in 3D and yields local coordinates that are 
independent of the environment's topology.  Thus cognitive demands and the range of animal's natural movement patterns are likely to influence the symmetries in the 
arrangement of grid fields.

Grid cells, which represent the position of an animal~\cite{Hafting2005} have been discovered only recently. 
By comparison, in technical systems,  it has  been known since the 1950's that  the optimal quantizers for 2D signals rely on hexagonal lattices~\cite{Gray1998}. 
In this context, we note that lattice codes are also ideally suited to cover spaces that involve sensory or cognitive variables other than location. In higher-dimensional 
feature spaces, 
the potential gain could be dramatic. For instance, the optimal eight dimensional lattice is about $16$ times denser 
than the orthogonal 8D lattice~\cite{Conway1992} and would, therefore,  dramatically
increase the resolution of the corresponding  population code.
Advances in experimental techniques, which allow one to simultaneously record from
large numbers of neurons~\cite{Ahrens2013,Deisseroth2013} and to automate stimulus delivery for dense parametric mapping~\cite{Scott2004}, now pave the way to 
search for such representations in cortex.  For instance, by parameterizing $19$ metric features of cartoon faces, such as hair length, iris size, or eye size, Freiwald et al. 
showed that face-selective cells are broad tuned to multiple feature dimensions~\cite{Freiwald2009}.  Such joint feature spaces should be the norm rather than the exception~\cite{Rigotti2013}.
While no evidence for lattice codes was found in the specific case of face-selective cells, data sets like this one will be the test-bed for our hypothesis.

\section*{Methods}

We study population codes of neurons encoding the $D$-dimensional space by considering 
the Fisher information $\bm J$ as a measure for their resolution.
The population coding model, the construction
to periodify a tuning shape $\Omega$ onto a lattice $\mathcal{L}$ with center density $\rho$, as 
well as the definition of the Fisher information, are given in the main text. 
In this section we give further background on the methods. 

\subsection*{Scaling of grid cells and the effect on $\bm J_{\varsigma}$}

How is the resolution of a grid module affected by dilations? Let us assume we 
have a grid code with signature $\varsigma=(\Omega,\rho,\mathcal{L})$, as 
defined in the main text, and that $\lambda > 0$ is a scaling factor. Then 
$\lambda \varsigma:=(\Omega(\lambda r),\rho(\lambda x),\lambda \cdot 
\mathcal{L})$ is a grid module, too, and the corresponding tuning curve 
${(\Omega\circ\lambda)}_{\lambda \mathcal{L}}$ satisfies:
\begin{equation}
 {(\Omega\circ\lambda)}_{\lambda \mathcal{L}}(x) = \Omega_\mathcal{L}(\lambda 
x).
\end{equation}
Thus, the tuning curve ${(\Omega\circ\lambda)}_{\lambda \mathcal{L}}$ is a 
scaled version of $\Omega_\mathcal{L}$. What is the relation between the Fisher 
information of the initial grid module and the rescaled version? Let us fix the 
notation: $\rho(c) = \sum_i^N \delta(c-c_i)$. From the definition of the population 
information (Eq.~\eqref{PopulationFisherinformationDef}),
we calculate
 \begin{equation}
 \bm J_{\lambda\varsigma} (0) = \sum_i \bm J_{{(\Omega\circ\lambda)}_{\lambda 
\mathcal{L}}}(\lambda c_i) = \sum_i \bm J_{{\Omega}_{\mathcal{L}}}(c_i) \cdot 
\frac{1}{\lambda^2} = \frac{1}{\lambda^2}  \bm J_{\varsigma} (0),
\end{equation}
where in the second step 
we used the re-parametrization formula of the Fisher 
information~\cite{Lehmann1998}.
This shows that the Fisher information of a grid 
module scaled by a factor $\lambda$ is the same as the Fisher information of the 
initial grid module times $1/\lambda^2$.

\subsection*{Population FI for Poisson noise with radially symmetric tuning}

In the results  section, we give a concrete example for Poisson noise and the bump function. Here we 
give the necessary background. Eq.~\ref{PopulationFisherinformation} states that 

\begin{equation}
\bm J_{\varsigma}(0) \approx \frac{M}{\mathrm{det}(\mathcal{L})}
\int_{B_R(0)}
\bm J_{\Omega^\mathcal{L}}(c) \mathrm{d}c. \nonumber \end{equation}
One would like to know $\int_{B_R(0)}
\bm J_{\Omega^\mathcal{L}}(c) \mathrm{d}c$ for various tuning shapes $\Omega$ with $\mathrm{supp}(\Omega) \leq R$. 
 
Consider $x \in L$ and $\alpha \in \{1, 
\ldots , D\}$. Then:
\begin{equation}
 \frac{\partial \ln P(K|x)}{\partial x_\alpha}  = \frac{\partial \ln P
(K,s)}{\partial s}
\Big|_{s=\Omega^\mathcal{L}(x)} \cdot \Omega'(\|x\|^2)  f_{max}\tau 2 x_\alpha .\label{individualpart}
\end{equation}
Together with the definition of the FI Eq.~\eqref{DefFI}, this 
yields
\begin{eqnarray}
 \bm {J_{{\Omega^\mathcal{L}}}}(x)_{\alpha \beta} & = & 4 x_\alpha x_\beta 
f_{max}^2 \tau^2
\Omega'(\| x\|^2)^2  \cdot \\ & & \underbrace{\sum_K
\left(\frac{\partial}{\partial s} \ln P
(K,s)\Big|_{s=\Omega^\mathcal{L}(x)}\right)^2 \cdot P
(K,\Omega^\mathcal{L}(x))}_{=:\mathcal{N}(\|x\|^2)}. \nonumber
\end{eqnarray}
Note that for $\alpha \neq \beta$ this function is odd in $x$. Thus, when
averaging these individual contributions over the symmetric fundamental domain 
$L$:
$\int_{L} {\bm {J}_{\Omega^\mathcal{L}}(c)}_{\alpha \beta}
\mathrm{d}c = 0$ for $\alpha \neq \beta$. Thus, the diagonal entries are all identical.

For Poisson spiking $\mathcal{N}(\|c\|^2)$ has a particularly simple
form, namely $\mathcal{N}(\|c\|^2) = 1/(f_{max}\tau \Omega(\|c\|^2))$. 
The trace of the Fisher information matrix becomes
\begin{equation}
\textrm{tr} \left( \bm J_{\varsigma}(0)  \right) =  4 f_{max}\tau \int_{B_R(0)}
\underbrace{ \|c\|^2 \frac{\Omega'( \|c\|^2 
)^2}{\Omega(\|c\|^2)}}_{=:\mathcal{F}(c)}
\textrm{d}c. \label{TraceofJ_POI}
\end{equation}
Thus, the trace only depends on the tuning shape $\Omega$ and its first 
derivative. In the main text, we use the following specific tuning shape:
\begin{equation}
 \Omega(r) =  \begin{cases}
               \exp 
\left(-\frac{\theta_1}{\theta_2^2-r^2}+\frac{\theta_1}{\theta_2^2} \right ) 
&\mbox{if} \ |r|<\theta_2 \\ 0 &\mbox{otherwise}
\end{cases} \label{TuningcurveBump}\end{equation}

\begin{figure}[ht!]
\includegraphics[angle=0,width=\columnwidth]{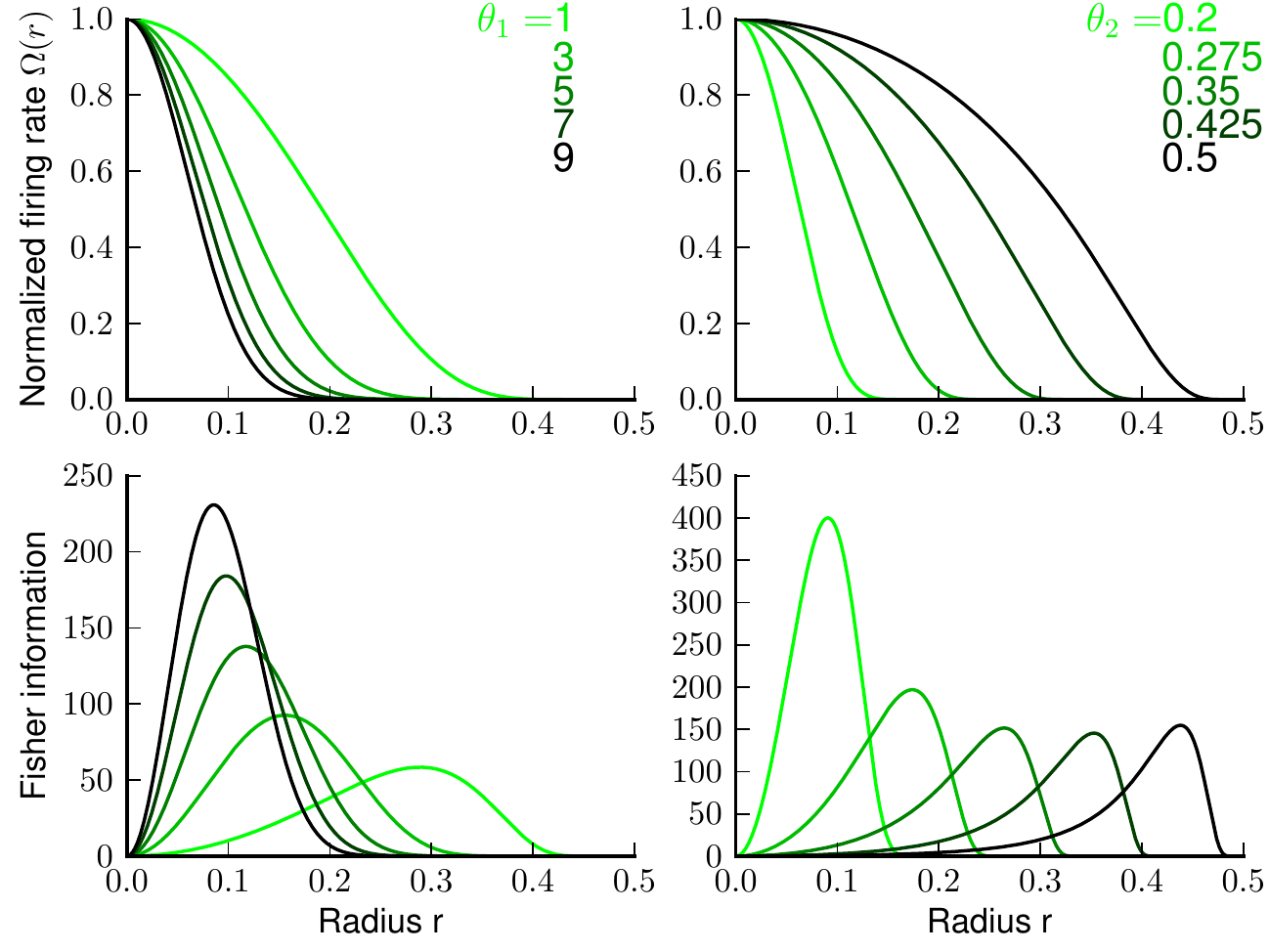} 
\caption{The firing rate and Fisher information of the bump  tuning shape. Upper left panel: 
Tuning shape $\Omega(r)$ with parameters $\theta_2=0.5$ and varying $\theta_1$. 
Lower left panel: Corresponding Fisher information $\mathcal{F}(r)$. Upper right 
panel:  Tuning shape $\Omega(r)$ with parameters $\theta_1=0.25$ and varying 
$\theta_2$. Lower right panel: Corresponding Fisher information 
$\mathcal{F}(r)$.}
\label{fig:IllustrationBump}
\end{figure}

This type of function is often called 'bump function' in topology, as it has 
a compact support but is everywhere smooth (i.e. infinitely times continuously differentiable). In particular, the support of this 
function is $[0,\theta_2)$, and is therefore controlled by the parameter 
$\theta_2$. The other parameter $\theta_1$ controls the slope of the bump's flanks (see 
upper panels of Fig.~\ref{fig:IllustrationBump}). 

For the bump-function $\Omega$ and radius $r=\sqrt{\sum_\alpha^D x_\alpha^2}$
the integrand for the FI is given by
\begin{equation}
\mathcal{F}(c) =  \begin{cases} \frac{4\theta_1^2 
r^2}{\left(\theta_2^2-r^2\right)^4}
               \exp 
\left(-\frac{\theta_1}{\theta_2^2-r^2}+\frac{\theta_1}{\theta_2^2} \right ) 
&\mbox{if} \ |r|<\theta_2 \\ 0 &\mbox{otherwise}
\end{cases} \label{J_TuningcurveBump}\end{equation}

The lower panels of Fig.~\ref{fig:IllustrationBump} depict the integrand of 
Eq.~\eqref{TraceofJ_POI}, defined as $\mathcal{F}(c)$. This functions shows 
``how much Fisher information particular cells at a particular distance 
contributes to the location $0$''. By integrating the FI over the fundamental domain $L$ for a lattice $\mathcal{L}$ one 
gets $\bm J_{\varsigma}(0)$, i.e. the average FI contributions from all neurons (as shown in Figures~\ref{fig:Results2D}, \ref{fig:Results3D} and \ref{fig:Results3Dpackings}E).

\begin{acknowledgments}

We thank Kenneth Blum for discussions. A.M. is grateful to Mackenzie Amoroso for 
graphics advice. This work was supported by the Federal Ministry for Education and Research 
(through the Bernstein Center for Computational Neuroscience Munich), by DFG grant MA 6176/1-1 
(A.M.) and the Marie Curie Fellowship Program of the European Union (A.M.). 
\end{acknowledgments}

\end{document}